# Metasurface spaceplates reach a millimeter-scale squeezed length of free space

Imon Kalyan[1], Raghvendra P. Chaudhary[1], and Nir Shitrit[1*]

[1]*School of Electrical and Computer Engineering, Ben-Gurion University of the Negev, Be'er Sheva 8410501, Israel*

(Dated: April 23, 2026)

[*]Corresponding author. E-mail: nshitrit@bgu.ac.il

**Abstract**

Metasurfaces offer compact flat lenses (metalenses) for miniaturized imaging systems; however, the utmost miniaturization requires not only metalenses but also a substantial reduction of free space. A Spaceplate is a flat-optics element designed to mimic free-space propagation, effectively propagating light over a distance far exceeding its physical thickness, with the induced squeezed length serving as the key figure of merit. Despite substantial progress, most existing spaceplate designs have been fundamentally constrained by a trade-off between squeezed length and numerical aperture, and none has demonstrated a feasible structure supporting both a moderate numerical aperture and a *millimeter*-scale squeezed length. Here, we report a metasurface spaceplate reaching the milestone of a millimeter-scale squeezed length with a practical numerical aperture. We achieved this by combining advantageous elements from existing approaches: high compression ratios and inverse-design flexibility in optimized multilayer metasurfaces—serving as the spaceplate unit structure—and preserving its numerical aperture by coupling its replicas—for constructing a coupled cascaded spaceplate with an increased thickness. For operation in the mid-wave infrared, we demonstrated an optimized spaceplate exhibiting a high compression ratio of ∼14 with a physical thickness of ∼80 μm—resulting in a squeezed length of 1.09 mm—for a numerical aperture of 0.13. We developed a general framework for calculating the transmission characteristics of multilayered spaceplates while optimizing their layer thicknesses to accurately reproduce the target free space. Strikingly, millimeter-scale squeezed lengths with practical numerical apertures via metasurface spaceplates pave the way for ultrathin imaging systems through their utmost miniaturization, opening a new paradigm for virtual and augmented reality headsets, cellphones, and many more.

## 1. Introduction

Flat optics has recently attracted great interest as a platform for miniaturizing optical systems by replacing conventional bulky optical components with metasurfaces [1–4]—engineered functional surfaces consisting of subwavelength scatterers. Metasurface-based flat optics achieves wavefront shaping at will through local control of phase, amplitude, and polarization in the spatial domain—by facilitating one-to-one correspondence between information pixels and subwavelength structures. These local metasurfaces enable compact (i.e., virtually flat, ultrathin, and lightweight) devices with unprecedented functionalities, including metalenses [5–8], beam deflectors [1,2,6,9], vortex generators [2,9–12], and holograms [13,14], among others, covering endless potential applications [15]. However, one essential optical component has long been overlooked in flat optics: free space (FS). FS is an integral part of most optical systems and typically constitutes a large portion of their overall volume—e.g., the object-to-lens and lens-to-image propagation distances in imaging systems. Therefore, achieving the utmost miniaturization of imaging systems requires not only compact flat lenses (metalenses) but also a substantial reduction of FS. FS fundamentally cannot be substituted with *local* flat optics characterized by *space*-dependent transfer functions, since FS propagation has a *momentum*-dependent transfer function [16,17]; in that regard, *nonlocal* flat optics [16–33] kicks in by offering to substitute FS with spaceplates [16,17,23–26,29,30]. Notably, local metasurfaces are inhomogeneous structures whose response is governed by an imprinted space-variant phase, while nonlocal metasurfaces are homogeneous structures whose response is controlled by the angle of incidence of light.

A spaceplate is a nonlocal flat-optics element engineered based on the target functionality of FS that effectively propagates light for a distance that can be considerably longer than the plate thickness [16,17,23–26,29,30]. A spaceplate with a physical thickness $d_{\mathrm{SP}}$ can replace an effective FS of length $d_{\mathrm{eff}} \gg d_{\mathrm{SP}}$ [see Fig. 1(a)], where the ratio between these two quantities $C = d_{\mathrm{eff}}/d_{\mathrm{SP}}$ is the compression factor of the spaceplate. While high values of compression factors are desired, application-wise, the foremost parameter for squeezing FS towards ultrathin imaging systems is the squeezed length induced by the spaceplate $D = d_{\mathrm{eff}} - d_{\mathrm{SP}} = (C - 1)d_{\mathrm{SP}}$. This spaceplate characteristic is proportional to both its thickness and compression ratio, thus indicating the length by which the spaceplate shortens the lens-to-focus (or lens-to-image) distance [see Fig. 1(a)]. Moreover, as nonlocal flat-optics elements, spaceplates operate in the momentum domain, so they are designed to support a numerical aperture (NA)—i.e., an angular range that sets the imaging resolution. To reach a large and impactful squeezed length $D$ (hundreds of wavelengths) by nonlocal optics, either the compression factor $C$ or the spaceplate thickness $d_{\mathrm{SP}}$, or both, should be maximized. For a $C$ slightly larger than 1, $d_{\mathrm{SP}}$ should be enlarged and unavoidably made thicker than $D$, as schematically illustrated in Fig. 1(b); this scenario fits spaceplate realizations based on unstructured natural birefringent materials (here, $C$ is the ratio of the extraordinary to the ordinary refractive index) [17]. On the other hand, by considering spaceplate realizations of resonant nonlocal flat-optics structures, such as photonic crystals [16] or multilayer metasurfaces with optimized subwavelength layers [17,24], $d_{\mathrm{SP}}$ is subwavelength or a few wavelengths; thus, a large $D$ necessitates an extremely high $C$ [see Fig. 1(b)]—which is achieved owing to the high Q-factors of such resonators, operating near their resonance frequencies. For example, photonic crystal slabs have demonstrated compression ratios exceeding 100 [16], while multilayer metasurfaces have reached more than 300 [24]. However, such single-resonator spaceplates [i.e., stand-alone resonators showing a single-mode operation; see Fig. 2(c) for a typical transmission response] exhibit a fundamental trade-off between compression ratio (or, alternatively, squeezed length) and NA [16,17,23,24,26], which hinders reaching the goal of large squeezed lengths for moderate NAs [see Fig. 1(c)]. This $D$–NA trade-off is a direct outcome of (i) an inherent mismatch between the typical S-shape (arctangent-like profile) of a resonator phase and the cosine profile of the FS phase—at large angles, and (ii) a limited phase range of maximum $\pi$ for single-resonator structures [23] [see Fig. 1(c) bottom inset].

To further increase the squeezed length, one may consider cascading multiple resonators, thereby increasing the total thickness of the composite spaceplate [16,23,30]. Chen *et al.* demonstrated that coupling subwavelength-thick Fabry–Pérot resonators to their replicas via quarter-wavelength layers allows the squeezed length of the cascaded structure to be increased without compromising the NA [23]. However, such layered Fabry–Pérot architectures are fundamentally limited in their achievable compression ratios: high values of $C \propto \frac{1+R}{1-R}$ require near-unity reflectivity $R$ of the reflecting layers; this, in turn, demands a large refractive-index contrast between the reflecting and cavity layers, as $R \propto (n - n_c)^2$, where $n$ and $n_c$ are the refractive indices of the reflecting and cavity layers, respectively [23,30]. Consequently, implementations based on common materials in the visible or infrared spectral regions yield only small compression ratios. In a recent study [30], an improved spaceplate design based on Fabry–Pérot resonators was demonstrated, achieving an enhanced $C$ by replacing the reflecting layers with high-reflectivity distributed Bragg reflectors; however, this design method is highly constrained: there is one-to-one correspondence between the predefined operating wavelength and layer materials and the resulting structure, from which the spaceplate characteristics ($C$, $D$, and NA) are directly determined [23,30]. This is in contrast to multilayer metasurfaces [17,24], where the structure is obtained by inverse design—optimizing the subwavelength layers to match the response corresponding to the target spaceplate characteristics—which, design-wise, provides substantially greater flexibility and scalability for engineering the spaceplate squeezed length. Strikingly, despite these advances, none of the design methods for spaceplates has thus far demonstrated a feasible structure that simultaneously supports a moderate NA and a *millimeter*-scale squeezed length.

Here, we report a spaceplate structure that reaches a millimeter-scale squeezed length with a practical NA. We achieved this milestone by combining the pros of high compression ratios and flexible and scalable design in multilayer metasurfaces with optimized subwavelength layers—serving as the spaceplate unit structure (SUS)—and preserving the NA of this SUS by coupling its replicas—for constructing a coupled cascaded spaceplate structure (CCSS) with an increased thickness. The obtained spaceplate overcomes the trade-off between squeezed length and NA while exhibiting a high compression ratio. For a mid-wave infrared wavelength, we obtained an optimized spaceplate exhibiting a compression ratio of $\sim$14 with a physical thickness of $\sim$80 μm—resulting in a squeezed length of 1.09 mm—for a NA of 0.13; the CCSS, exhibiting high average transmittance of 0.80, consists of only 8 replicas of the SUS. The multilayered spaceplate with alternating polycrystalline silicon and silicon dioxide subwavelength layers is a feasible structure benefiting from a mature technology for producing these stacks. We developed a methodology for calculating the transmission characteristics of multilayered spaceplates while optimizing their layer thicknesses to match the target of an equivalent FS. Moreover, we directly observed the key parameter of the squeezed length via a combined lens–spaceplate optical system; by introducing the spaceplate, the position of the focal plane is shifted towards the lens—manifesting the FS squeezing by the spaceplate. Strikingly, reaching the milestone of millimeter-scale squeezed length with a practical NA by metasurface spaceplates paves the way towards ultrathin imaging systems; moreover, the combination of local and nonlocal flat optics—e.g., metalens–spaceplate integration—provides the route for the utmost miniaturization of optical systems, leveraging solar concentrators, spectrometers, virtual and augmented reality headsets, cellphones, and many more.

## 2. Methodology

A monochromatic optical field propagating in FS—a space-invariant linear system—is characterized by a momentum-dependent transfer function [16,17]

$$H(k_x,k_y) = \exp[ik_0 d\cos\theta(k_x,k_y)], \quad (1)$$

where $k_0 = 2\pi/\lambda$ is the FS wave number, $\lambda$ is the FS wavelength, $d$ is the distance between the input and output planes, and $\cos\theta = \frac{1}{k_0}\sqrt{k_0^2 - k_\parallel^2}$, where $k_\parallel = \sqrt{k_x^2 + k_y^2}$ is the in-plane wave number (i.e., the projection of the wave vector in the $x$–$y$ plane). We consider a propagating plane wave with $k_\parallel < k_0$. In this formalism, $\theta$ is the angle of incidence on the spaceplate. The light propagating through a spaceplate with a thickness $d_{\mathrm{SP}}$ acquires a phase $\phi_{\mathrm{SP}}$ that mimics the phase pickup by propagating in FS a distance $d_{\mathrm{eff}} = Cd_{\mathrm{SP}}$, expressed as $\phi_{\mathrm{FS}} = k_0 C d_{\mathrm{SP}}\cos\theta$—which is the target phase. The NA of the spaceplate corresponds to the maximum angle for which the spaceplate phase closely matches the target FS phase with a high Strehl ratio $S = \exp(-\sigma^2)$, where $\sigma^2 = \langle(\phi_{\mathrm{FS}} - \phi_{\mathrm{SP}})^2\rangle_\theta$ (i.e., the average over the angle range of the difference squared between the FS and spaceplate phases) [24]. To achieve a spaceplate phase $\phi_{\mathrm{SP}}$ that mimics the FS phase $\phi_{\mathrm{FS}}$ with targeted compression factor $C$ and NA, we first construct a SUS of a multilayered single resonator comprising a fixed odd number of alternating polycrystalline silicon (poly-Si) and silicon dioxide ($SiO_2$) subwavelength layers [see Fig. 2(a)]. At the wavelength of 4.05 μm, we considered the refractive indices of 3.594 and 1.387 for poly-Si and $SiO_2$, respectively. Initially, we assign a random thickness to each layer in the multilayered structure, within the range of 0.1 μm to 1.5 μm, while setting the total thickness of the multilayers to $d_{\mathrm{SP}}{\sim}10$ μm. Then, we calculate the transmittance ($T$) and phase ($\phi_{\mathrm{SP}}$) of the multilayered structure as a function of the angle of incidence, within the angular range of $\theta = 0°$ to 7.5°, corresponding to a target NA of 0.13 [see Fig. 2(c)] (the initially considered NA is chosen to be slightly larger than the target NA to avoid reduced transmission near the operational limit). The calculation is performed for a transverse magnetic polarization, which exhibits higher compression factors compared to a transverse electric polarization [24]. We developed an optimization algorithm based on the gradient descent method [34] that minimizes $\sigma^2$ (which represents the difference between the FS and spaceplate phases) [see Fig. 2(e)] by varying the thicknesses of the layers. In this optimization method, $C$ of the final optimized multilayered structure is typically slightly different from the target value [24]. Therefore, we extract $C$ by fitting linearly the calculated spaceplate phase $\phi_{\mathrm{SP}}$ versus $\cos\theta$ and dividing the obtained slope by $k_0 d_{\mathrm{SP}}$. Structures with a Strehl ratio $S < 0.85$ are rejected from a further assessment. To ensure high transmittance of the optimized structure, the average transmittance $T_{\mathrm{avg}}$ is assessed; if $T_{\mathrm{avg}} < 0.5$, the optimization is terminated, and a new trial is initiated using a new set of randomly generated thicknesses. In the context of spaceplate designs, this optimization procedure allows us to obtain optimized multilayered single-resonator metastructures that outperform Fabry–Pérot resonators by exhibiting similar transmission characteristics—but with significantly higher compression factors. We refer to this optimized multilayered single-resonator structure as the SUS.

Transmission calculations of a SUS with subwavelength layers can be performed efficiently and rapidly by employing the powerful transfer matrix method (TMM) [35,36]. Figure S1 shows transmission calculations of an optimized multilayered structure obtained using the quasi-analytical TMM; to compare, we also calculated the same transmission characteristics of this structure using commercial solvers (COMSOL Multiphysics and PlanOpSim). The transmission calculations of the TMM and the two commercial solvers (based on a finite element method or a rigorous coupled-wave analysis) exhibit excellent agreement (see Supplement 1, Sec. 2). While all methods show a similar high accuracy, the TMM is significantly faster. Therefore, we optimize the multilayered SUS by employing the TMM.

As noted in Ref. [24] and demonstrated in Fig. 1(c), an optimized multilayered single-resonator structure can reach a large squeezed length by increasing the compression factor—but with the significant cost of substantial reduction in the NA. Figure 1(c) shows that an optimized multilayered single-resonator SUS with a thickness of $d_{\mathrm{SP}}{\sim}11$ μm can reach a squeezed length of $D{\sim}1$ mm by obtaining a high compression ratio of $C{\sim}92$; however, such an optimized multilayered single-resonator spaceplate exhibits an extremely low NA of 0.04 [see the green dot in Fig. 1(c)]. This behavior arises from two fundamental factors: (i) the inherent mismatch between the typical S-shape (arctangent-like) phase response of a resonator and the cosine dependence of the FS phase at large incidence angles, and (ii) the intrinsically limited phase range in single-resonator structures. To address this challenge, we construct a CCSS consisting of multiple repetitions of the achieved optimized SUS separated by a $SiO_2$ spacer layer. Similar to the SUS, we calculate the transmittance and phase of the cascaded structure as a function of the angle of incidence using the TMM; the compression factor $C$ and the resulting squeezed length $D$ are extracted by fitting the calculated phase to the phase of the equivalent FS. Note that the compression factor of the CCSS can differ slightly from that of the SUS [see Fig. 1(d) inset]. Here, unlike Fabry–Pérot resonators, the thicknesses of the spacer layers (i.e., the distance between two neighboring SUSs) are not fixed and require further optimization. This stage is crucial, as without a proper optimization of the spacer layer thicknesses, the phase of the CCSS can deviate significantly from the phase of the equivalent FS, resulting in a high $\sigma^2$ (undesirable low Strehl ratio). Moreover, the transmission characteristics of the CCSS are highly sensitive to the thickness of each spacer layer, such that mismatched thicknesses can give rise to low transmittance within the target angular range. To obtain a spaceplate that closely matches the equivalent FS response, we optimized the thickness of each spacer layer by simultaneously maximizing $S$ and the minimum transmittance $T_{\mathrm{min}}$ for the target unit-structure NA. We first assigned uniform thicknesses to all spacer layers and varied them collectively to minimize $\sigma^2$. Afterwards, we fine-tuned the thickness of each spacer layer independently to further increase $T_{\mathrm{min}}$, while maintaining $S$ within 3% of its maximum value (see Supplement 1, Sec. 3 for more details). By employing this coupling optimization procedure, a squeezed length of $D\sim 1$ mm is achieved by a CCSS consisting of only 8 replicas of the optimized SUS, as shown in Fig. 1(d). Note that a thick optimized multilayered single-resonator SUS structure with a similar thickness and compression ratio compared to the CCSS can exhibit a similar squeezed length—but with an extremely low NA [see the square point in Fig. 1(d)]. The angle-dependent phase response of a single-resonator spaceplate, after subtracting a global angle-independent phase, is bounded by $-\pi < \phi(\theta) < 0$ [23,30] [see Fig. 1(e)]. This bound constrains the maximum angle—i.e., the NA—over which the spaceplate can accurately mimic FS propagation, such that $\phi_{\mathrm{FS}}(\theta_{\mathrm{max}}) - \phi_{\mathrm{FS}}(0) = k_0 C d_{SP}(\cos\theta_{\mathrm{max}} - 1) = -\pi$. This relation yields an upper limit on the unit-structure NA as $\mathrm{NA}\sim\sqrt{\lambda/(Cd_{\mathrm{SP}})}$ [23,30]. Note that by coupling $N$ replicas of the SUS, the phase is bounded by $-N\pi$ [23,30] [see Fig. 1(e)], while the cascading scales the thickness roughly to $Nd_{\mathrm{SP}}$ (here, $d_{\mathrm{SP}}$ is the SUS thickness). Therefore, the CCSS obeys the same limit, thereby preserving the unit-structure NA [Fig. 1(e)]—while increasing the squeezed length [Fig. 1(d)].

## 3. Results

Figure 2(a) depicts the optimized multilayered single-resonator structure serving as the SUS. This optimized SUS consists of 13 poly-Si–$SiO_2$ alternating subwavelength layers with a total thickness of 9.74 μm, where the optimized thicknesses are within the range of 0.20 μm to 1.29 μm (see Supplement 1, Sec. 1 for the detailed layer thicknesses). Figure 2(c) shows the dependence of the transmittance and the transmitted phase on the incident angle for the SUS. Note that the plotted phase is offset by a global angle-independent phase, such that the phase at $\theta = 0°$ equals zero. This optimized design exhibits a compression factor of 13.30 (the target compression factor was 13), giving rise to a squeezed length of 0.12 mm within the target angular range of $\theta = 0°$ to 7.5° (NA = 0.13), at the working wavelength of 4.05 μm. Note that the optimization conditions are satisfied with a near-perfect Strehl ratio of $S = 0.99$ and an average transmittance of $T_{\mathrm{avg}} = 0.65$. Figure 2(e) presents $\sigma^2$ (the average over the angle range of the difference squared between the FS and spaceplate phases—the cost function minimized by the optimization algorithm) as a function of the iteration number throughout the optimization procedure of the SUS. The plot reveals the successful optimization manifested by a monotonic decrease of $\sigma^2$ with more than 100-fold reduction. Figure 2(b) depicts the optimized CCSS constructed by 8 replicas of the SUS (111 alternating layers) with a total spaceplate thickness of $d_{\mathrm{SP}} = 82.47$ μm (see Supplement 1, Sec. 1 for the detailed layer thicknesses). The corresponding transmittance and transmitted phase as a function of the incident angle are shown in Fig. 2(d). This optimized spaceplate supporting a NA of 0.13 exhibits a compression factor of 14.23, giving rise to a squeezed length of 1.09 mm ($\sim 270\lambda$ in wavelength units); the fitting of the spaceplate phase to the FS phase also yields a Strehl ratio of $S = 0.89$, and the average transmittance of this spaceplate is $T_{\mathrm{avg}} = 0.80$. Note the extended phase range of $\sim 5\pi$ associated with the CCSS [Fig. 2(d)] compared to the phase range of $\sim 0.55\pi$ associated with the SUS [Fig. 2(c)], which is the signature of coupling. Figure 2(f) presents the iterative evolution towards reaching the optimized CCSS, where the minimum transmittance and the Strehl ratio are shown as a function of the iteration number; we aim to maximize both parameters by optimizing the thickness of each spacer layer. The spaceplate parameters of the SUS and the CCSS are summarized in the table of Fig. 2(g). The obtained optimized spaceplate reaches the milestone of millimeter-scale squeezed length with a practical NA, thus paving the way towards ultrathin imaging systems.

We further verified the squeezed length of the optimized spaceplate by directly observing this quantity in a combined lens–spaceplate optical system. The role of the lens is to focus a monochromatic normally incident collimated beam, where the angles introduced by the focused light mimic the angle of incidence on the spaceplate—placed just after the lens. In this configuration, the NA (angle range) of the lens and the NA of the spaceplate are matched. By introducing the spaceplate, the position of the focal plane is shifted towards the lens—which is the manifestation of squeezing FS by the spaceplate. By comparing the positions of the focal plane with and without the spaceplate, we directly extract the observable quantity of the squeezed length $D$. We performed this procedure both theoretically and numerically. By considering a monochromatic, unit-amplitude, normally incident plane-wave illumination, we theoretically expressed the electric scalar field at the lens plane (just after the lens) as

$$E_1(x,z=0)=\exp[i\phi(x)], \tag{2}$$

where $\phi(x)=-k_0\left(\sqrt{x^2+f^2}-f\right)$ is the phase profile of the converging lens and $f$ is its focal length; for simplicity, we considered a one-dimensional (cylindrical) lens. Then, the Fourier transform of $E_1(x,z=0)$ is given by $\mathcal{E}_1(k_x,z=0)=\mathcal{F}\{E_1(x,z=0)\}$, where $k_x$ is the in-plane wave number—the projection of the wave vector along the $x$ axis. Based on the momentum-dependent (nonlocal) transfer function of FS [Eq. (1)], we expressed the transfer function of the spaceplate as

$$H_{\mathrm{SP}}(k_x)=\exp\left(iCd_{\mathrm{SP}}\sqrt{k_0^2-k_x^2}\right), \tag{3}$$

where $C$ and $d_{\mathrm{SP}}$ are the compression factor and the thickness of the obtained optimized spaceplate, respectively. In the momentum space, the transmission through the spaceplate is represented by multiplying by its transfer function, such that the field just after the spaceplate (i.e., the lens–spaceplate system) is given by

$$E_2(x)=\mathcal{F}^{-1}\{\mathcal{E}_1(k_x,z=0)H_{\mathrm{SP}}(k_x)\}, \tag{4}$$

where $\mathcal{F}^{-1}$ is the inverse Fourier transform. Note that in this theoretical analysis, the lens and the spaceplate are considered infinitesimally thin elements. Finally, by employing the Fresnel diffraction integral [37], the field at an observation axis $(x';z)$ in FS is calculated as

$$E_3(x';z) = \frac{\exp(ik_0 z)}{\sqrt{i\lambda z}} \int E_2(x) \exp\left[i \frac{k_0}{2z}(x' - x)^2\right] dx, \tag{5}$$

where $z$ is the distance between the lens–spaceplate plane and the observation plane. By calculating $E_3(x';z)$ for different propagation distances $z$, we obtained the intensity distribution of light propagating in FS after passing through the combined lens–spaceplate system. We considered a lens with a diameter of 0.8 mm and a focal length of 3.04 mm (the NA of the lens is ~0.13 which matches the NA of the spaceplate), at the wavelength of 4.05 μm. The results of this theoretical model are presented in Figs. 3(a)–3(c), where the intensity distributions for a stand-alone lens, the combined systems of a lens and the optimized SUS, and a lens and the optimized CCSS, are shown, respectively. The shift of the focal plane towards the lens with respect to the reference case of a stand-alone lens [Fig. 3(a)] reveals the squeezed length induced by the spaceplate; particularly, the observable quantity of the squeezed length corresponding to the optimized SUS is 0.115 mm [Fig. 3(b)], while the squeezed length corresponding to the optimized CCSS is 1.092 mm [Fig. 3(c)]. These theoretical results obtained by observing the focal shift in the lens–spaceplate system exhibit excellent agreement with the numerical results obtained by fitting the spaceplate phase to the target FS phase.

The squeezed length observed via the lens–spaceplate system was also characterized numerically, where an actual plano-convex lens (with same parameters as in the theoretical calculations) and the optimized multilayered spaceplate are considered. Here, the field just after the lens–spaceplate system $E_2(x)$ is calculated by a full-wave simulation (COMSOL Multiphysics), and the intensity distribution in the $x'$–$z$ plane is calculated by employing the Fresnel diffraction integral [Eq. (5)]. The observable squeezed lengths induced by the spaceplate corresponding to the optimized SUS and the optimized CCSS are 0.122 mm [Fig. 3(e)] and 1.015 mm [Fig. 3(f)], respectively [for comparison, see Fig. 3(d) for the corresponding reference case of a stand-alone lens]. We also present the intensity cross sections at the focal plane [Figs. 3(g)–3(i)]. While the intensity cross sections of the stand-alone lens show a full width at half maximum (FWHM) of 0.14 mm [Fig. 3(g)], exhibiting good agreement with the diffraction-limited value of $\lambda/(2\text{NA})$ ~0.15 mm, the cross sections corresponding to the lens–spaceplate systems show identical FWHM values [Figs. 3(h) and 3(i)]; this observation reveals that the spaceplate squeezes FS by shifting the focal plane towards the lens—without affecting the focusing or imaging characteristics (spot size, NA, magnification, etc.). Figure 3(j) shows the intensity cross sections along the optical axis which clearly reveal the squeezed length of $D \sim 1$ mm induced by the optimized CCSS. These results exhibit good agreement with the theoretical and numerical observations, concluding that the reported spaceplate reaches the milestone of millimeter-scale squeezed length with a practical NA. Note that the squeezed length can be further increased by realizing multiple copies $m$ of the spaceplate and then cascading them, such that the squeezed length is $m$-fold magnified.

## 4. Discussion and summary

The reported metasurface spaceplate, achieving a millimeter-scale squeezed length of FS while supporting a practical NA, combines advantageous elements from two complementary design approaches: multilayer metasurfaces with optimized subwavelength layers [17,24] and coupled cascaded Fabry–Pérot structures [23,30]. The first method benefits from high compression ratios and flexible, scalable inverse design but suffers from the trade-off between squeezed length and NA in single-resonator spaceplates. On the other hand, the second approach benefits from maintaining the NA of the single Fabry–Pérot resonator while coupling its replicas in a cascaded structure, thereby increasing the spaceplate thickness; however, the achievable compression ratios of Fabry–Pérot resonators are limited. We achieved millimeter-scale squeezed length with a moderate NA by combining the pros of high compression ratios in multilayer metasurfaces—serving as the SUS—and preserving the unit-structure NA by coupling its replicas—for constructing a CCSS with an increased thickness. Other approaches targeted large-scale squeezed length—but encountered major roadblocks. For example, one of the realizations of spaceplates is a uniaxial crystal, where the momentum-dependent transfer function is achieved by the dependence of the refractive index on the incident angle in birefringent materials [17]. In this realization, the compression factor is the ratio of the extraordinary to the ordinary refractive index, yielding a compression ratio slightly greater than 1; a squeezed length on the order of millimeters was attainable due to the unavoidable centimeter-scale thickness of the crystal [17]. Moreover, the space-squeezing experiment was conducted in an oil medium, as the refractive index of the surrounding medium must match the extraordinary refractive index of the bulky, thick uniaxial crystal [17]. These issues pose significant limitations on the realization of spaceplates using birefringent materials. Another demonstration of space squeezing used a spaceplate with a refractive index lower than that of the surrounding medium [17]—a fundamental challenge for FS squeezing. Additionally, Sorensen *et al.* demonstrated a three-lens spaceplate manifested by a large spaceplate thickness (∼300 mm) and a high compression ratio (up to 300) [29]; while this design indicates a large-scale squeezing of FS, its NA is extremely low (0.035)—rendering spaceplate designs based on lenses impractical. As for spaceplate realizations based on Fabry–Pérot resonators, they have been experimentally demonstrated using reflective metasurfaces in ambient air at microwave frequencies [26]; however, implementing this design at visible or infrared wavelengths with suspended subwavelength-spaced structures is impractical. In a similar fashion, Guo *et al.* considered cascading multiple photonic crystal slabs with subwavelength air gaps between adjacent resonators, thereby increasing the total thickness of the composite spaceplate to further increase the squeezed length [16]; nevertheless, this approach is impractical as realizing such suspended subwavelength-spaced multiple structures remains an arduous fabrication task. Notably, to reach a large and impactful squeezed length, either the compression factor or the spaceplate thickness, or both, should reach their upper limits, which are set by physical or technological considerations, respectively.

In summary, we report a metasurface spaceplate that reaches the milestone of a millimeter-scale squeezed length while supporting a practical NA. We achieved this large-scale compression of FS by combining two complementary design strategies: (i) high compression ratios and flexible, scalable inverse design in multilayer metasurfaces with optimized subwavelength layers—serving as the SUS—and (ii) preserving the unit-structure NA by coupling its replicas—for constructing a CCSS with an increased thickness. The obtained spaceplate overcomes the trade-off between squeezed length and NA while sustaining a high compression ratio. For operation in the mid-wave infrared, we demonstrated an optimized spaceplate with a compression ratio of ∼14 and a physical thickness of ∼80 μm—corresponding to a squeezed length of 1.09 mm—for a NA of 0.13; the CCSS, exhibiting a high average transmittance of 0.80, consists of only 8 replicas of the SUS. Importantly, the multilayered spaceplate—composed of alternating poly-Si and $SiO_2$ subwavelength layers—is a feasible structure benefiting from a mature technology for producing these stacks; particularly, the reported 80-μm-thick multilayered spaceplate is well matched to the fabrication technology of inductively coupled plasma chemical vapor deposition—a thin-film deposition technology producing high-quality, highly uniform films—thereby paving the way for future experiments. The reported spaceplate design can be achieved in the spectral region of interest; we targeted the mid-wave infrared, where system miniaturization by FS squeezing can, e.g., dramatically improve the thermal and mechanical stability of cryogenically cooled infrared detectors with integrated optics [38]. Strikingly, achieving a millimeter-scale squeezed length with a practical NA establishes metasurface spaceplates as a viable route towards ultrathin imaging systems. More broadly, the combination of local and nonlocal flat optics—such as metalens–spaceplate integration—opens a new paradigm for the ultimate miniaturization of optical systems, with potential impact on solar concentrators, spectrometers, virtual and augmented reality headsets, cellphones, and many more.

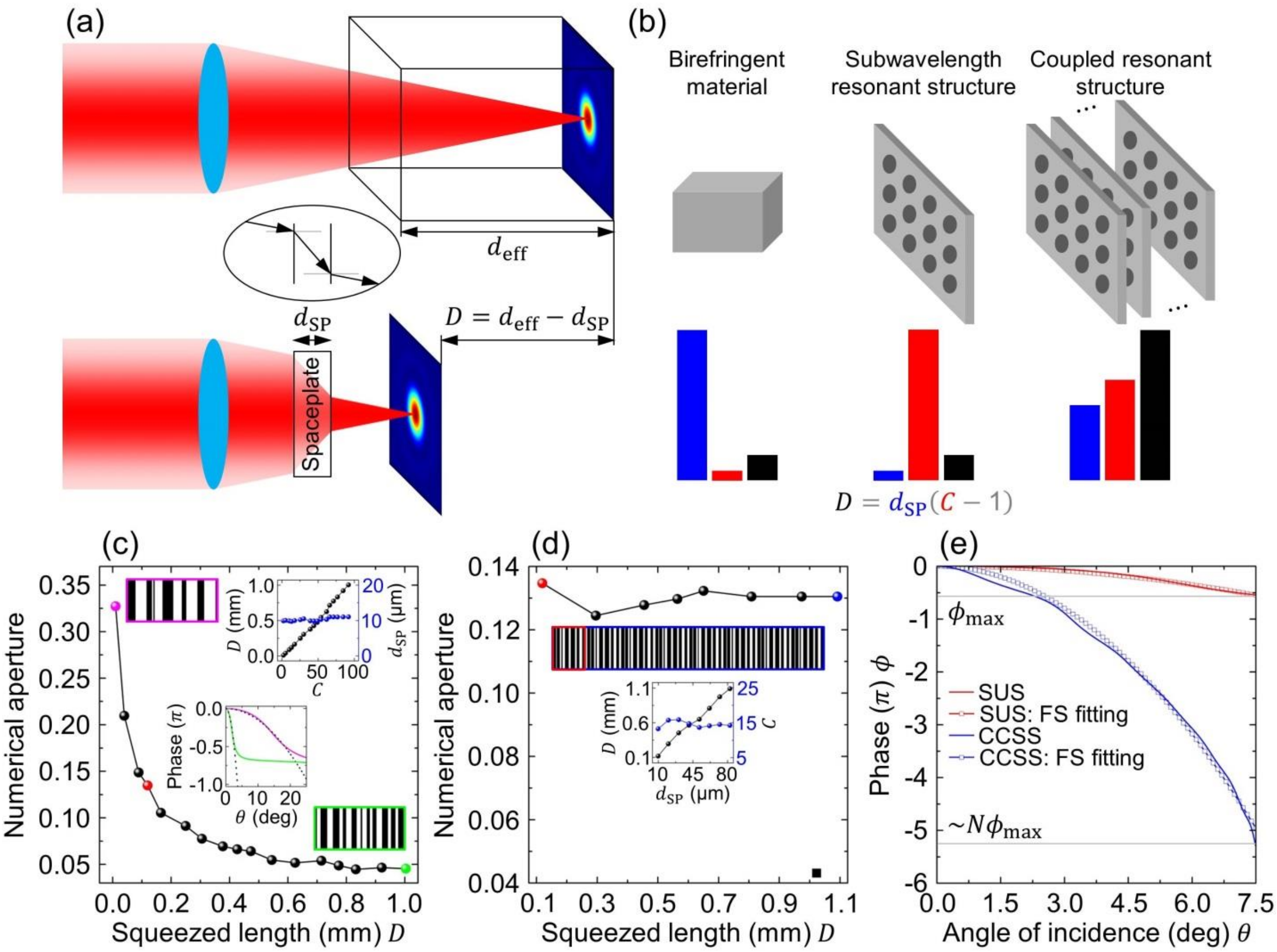

**Fig. 1.** Towards a large-scale squeezed length with a moderate numerical aperture: overcoming the trade-off between squeezed length and numerical aperture by coupled cascaded spaceplate structures. (a) Schematics showing the working principle of a FS squeezing by a spaceplate with its characteristics of the physical thickness $d_{\mathrm{SP}}$, the effective FS distance $d_{\mathrm{eff}}$, and the squeezed length $D$. The inset shows a typical refraction in the spaceplate with a refraction angle larger than the incident angle (i.e., the effective refractive index of the spaceplate is smaller than 1)—enabling the FS squeezing. (b) Schematic representation of the approaches for spaceplate realizations aiming to maximize the squeezed length via the relation $D = (C-1)d_{\mathrm{SP}}$. (c) Dependence of the NA on the squeezed length for different SUSs, revealing the trade-off between these parameters in single-resonator spaceplates. Each point in the plot corresponds to an optimized multilayer spaceplate (see the insets of the structures). The top inset shows $D$ and $d_{\mathrm{SP}}$ as a function of $C$, where $d_{\mathrm{SP}}$ is nearly constant. The bottom inset shows the dependence of the transmitted phase on the angle of incidence for the spaceplates with the smallest and largest $D$, illustrating the fundamental $D$–NA trade-off (the dotted lines represent the corresponding FS

phases). (d) Dependence of the NA on the squeezed length for the CCSS with a varying number of SUS replicas; here, the NA remains nearly constant. The bottom inset shows $D$ and $d_{\mathrm{SP}}$ as functions of $C$; here, $C$ is roughly constant, and $D$ increases as the number of coupled units $N$ is increased. The top inset shows the CCSS constructed from $N = 8$ replicas of the SUS. The square point shows the NA corresponding to a thick SUS with a thickness similar to that of the CCSS with 8 replicas; although they exhibit similar $D\sim1$ mm, the NA of the SUS is significantly lower. (e) Transmitted phase as a function of the angle of incidence for the SUS and the CCSS, revealing the extended phase range for the CCSS—which allows maintaining the unit-structure NA.

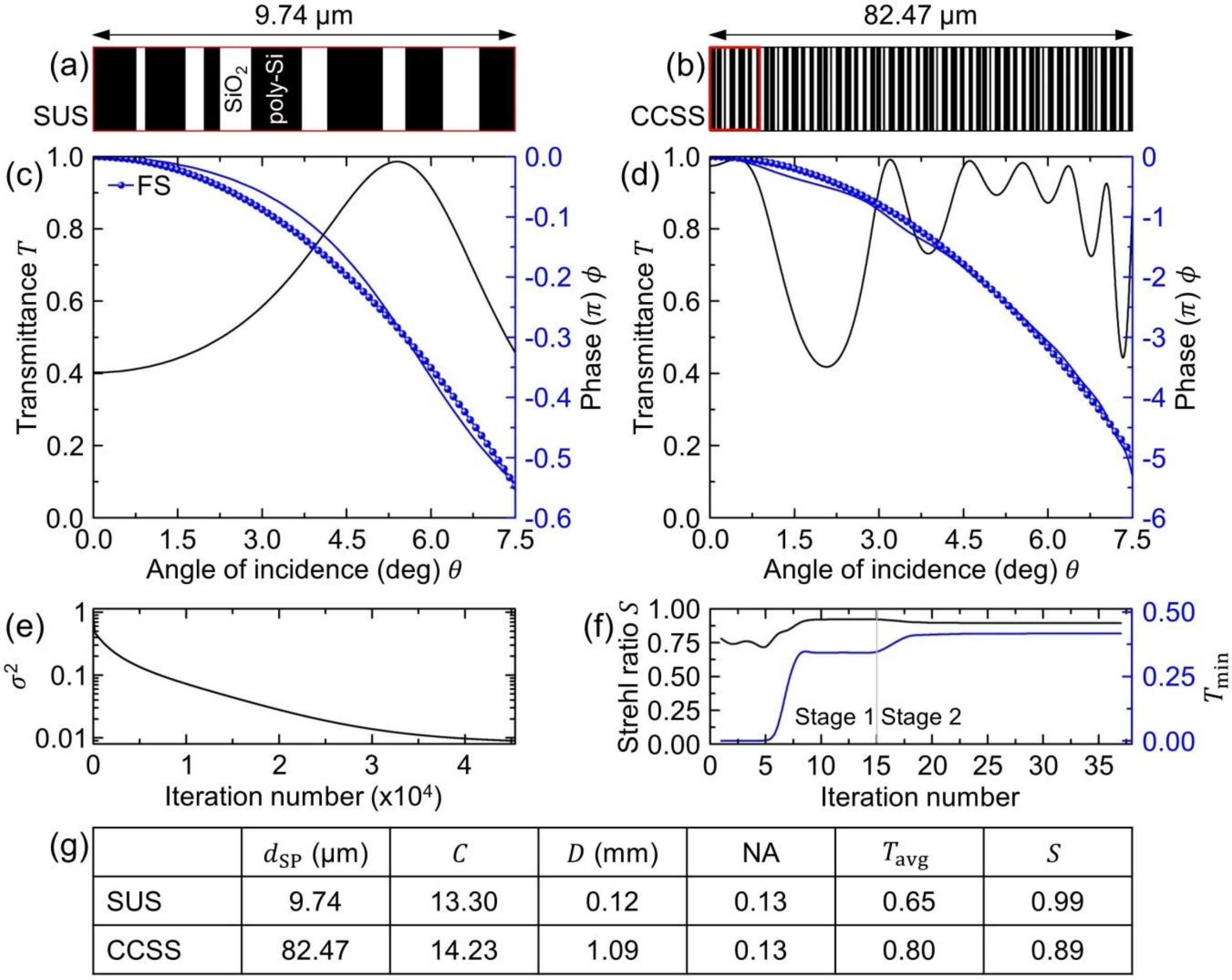

| | $d_{\mathrm{SP}}$ (μm) | $C$ | $D$ (mm) | NA | $T_{\mathrm{avg}}$ | $S$ |
|---|---|---|---|---|---|---|
| SUS | 9.74 | 13.30 | 0.12 | 0.13 | 0.65 | 0.99 |
| CCSS | 82.47 | 14.23 | 1.09 | 0.13 | 0.80 | 0.89 |

**Fig. 2.** Multilayered spaceplates: optimized unit structure and coupled cascaded structure. (a), (b) Schematic images of the optimized multilayered SUS ($d_{\mathrm{SP}} = 9.74$ μm, 13 layers) and the CCSS ($d_{\mathrm{SP}} = 82.47$ μm, 111 layers), respectively. The spaceplates are composed of alternating poly-Si and $SiO_2$ subwavelength layers with optimized thicknesses. The CCSS is constructed by 8 replicas of the SUS. (c), (d) Dependence of the transmittance and phase on the angle of incidence for the optimized SUS and the CCSS, respectively. The angular range from 0° to 7.5° corresponds to a NA of 0.13. The fitted phase of the equivalent FS is shown by the dots. For the SUS, $C = 13.30$ and $D = 0.12$ mm, whereas $C = 14.23$ and $D = 1.09$ mm for the CCSS. (e) $\sigma^2$ as a function of the iteration number, illustrating the cost function minimization throughout the optimization procedure of the SUS. (f) Strehl ratio and the minimum transmittance $T_{\mathrm{min}}$ as a function of the iteration number, presenting the evolution towards reaching the CCSS via the two-stage optimization. (g) Table denoting the characteristic spaceplate parameters for the SUS and the CCSS.

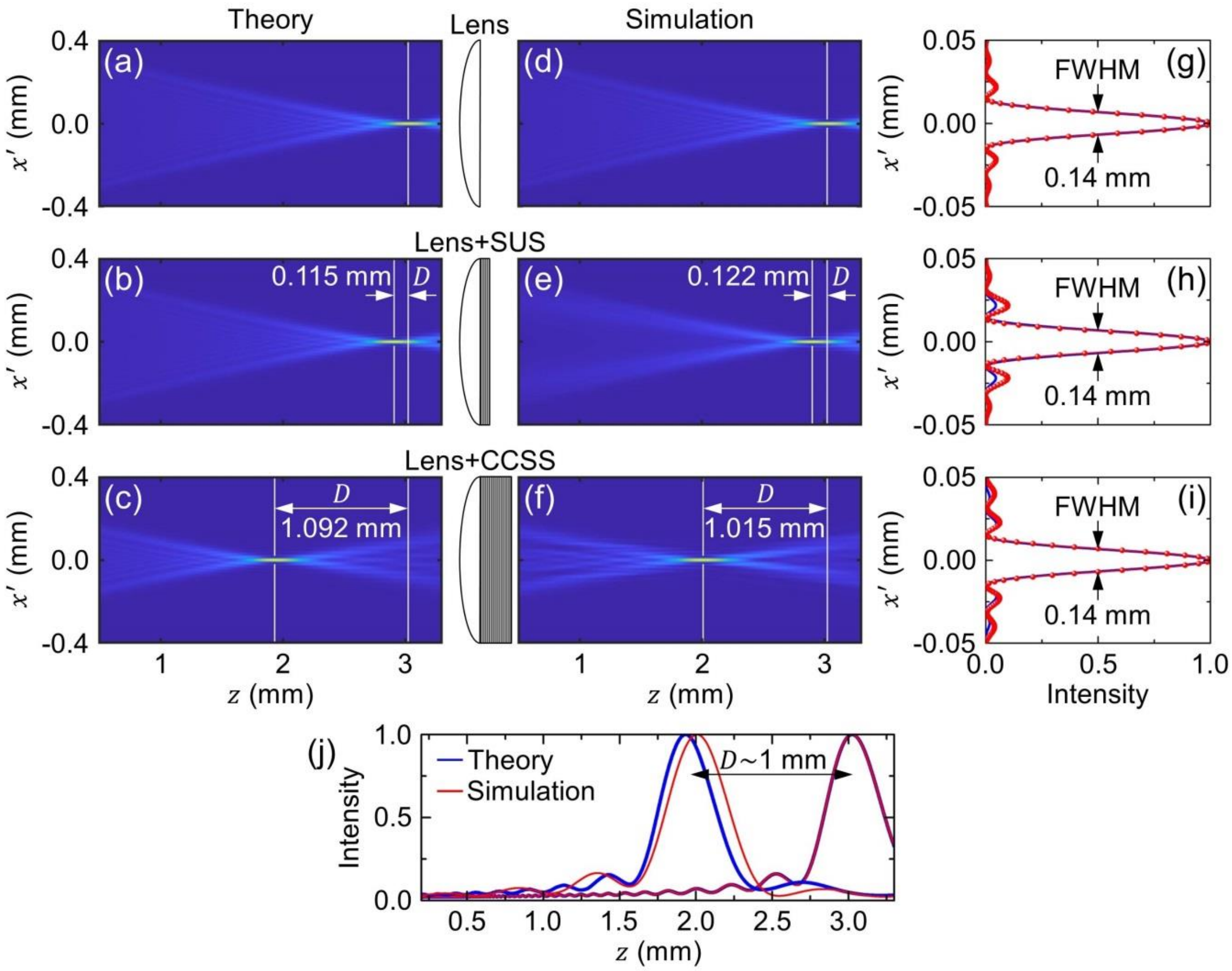

**Fig. 3.** Direct observation of the squeezed length by a lens–spaceplate system. (a) Theoretical calculation of the intensity distribution ($x'$–$z$ plane) showing the focusing by the stand-alone lens (the focal length is 3.04 mm). (b) Theoretical calculation of the intensity distribution of the lens–spaceplate system showing the focal plane shift towards the lens, i.e., the squeezed length induced by the SUS ($D = 0.115$ mm). (c) Theoretical calculation of the intensity distribution of the lens–spaceplate system showing the millimeter-scale squeezed length induced by the CCSS ($D = 1.092$ mm). (d)–(f) Numerical calculations of the intensity distributions of the stand-alone lens, lens–spaceplate systems with the SUS ($D = 0.122$ mm) and the CCSS ($D = 1.015$ mm), respectively. (g)–(i) Intensity cross sections [theory (blue) and numerical simulations (red)] at the focal plane of the stand-alone lens, the shifted focal planes of the lens–spaceplate systems corresponding to the SUS and the CCSS, respectively. Note the identical FWHM values expected from the spaceplate response. (j) Intensity cross sections (theory and simulations) along the optical axis clearly demonstrating the squeezed length of $D{\sim}1$ mm induced by the CCSS.

**Acknowledgments**

The authors gratefully acknowledge funding from the Israel Innovation Authority through its Metamaterials Consortium and the Israel Science Foundation (ISF) under Grant No. 1785/22.

**Disclosures**

The authors declare no conflicts of interest.

**Data availability**

Data underlying the results presented in this paper are not publicly available at this time but may be obtained from the authors upon reasonable request.

**Supplemental document**

See Supplement 1 for supporting content. Supplement 1 includes the layer thicknesses of the optimized multilayered spaceplate, the comparison between the transfer matrix method and commercial electromagnetic solvers, and the optimization method of the coupled cascaded spaceplate structure.

Supplemental Document for

# Metasurface spaceplates reach a millimeter-scale squeezed length of free space

Imon Kalyan[1], Raghvendra P. Chaudhary[1], and Nir Shitrit[1*]

[1]*School of Electrical and Computer Engineering, Ben-Gurion University of the Negev, Be'er Sheva 8410501, Israel*

[*]Corresponding author. E-mail: nshitrit@bgu.ac.il

## 1. Layer thicknesses of the optimized multilayered spaceplate

The layer thicknesses of the optimized multilayered spaceplate unit structure (SUS) are listed in Table S1. The optimized SUS consists of 13 alternating polycrystalline silicon (poly-Si) and silicon dioxide ($SiO_2$) subwavelength layers with a total thickness of 9.7406 μm. We construct a coupled cascaded spaceplate structure (CCSS) consisting of multiple repetitions of the optimized SUS, separated by a $SiO_2$ spacer layer. The optimized CCSS is constructed from 8 replicas of the SUS, consisting of 111 alternating layers, with a total spaceplate thickness of $d_{\mathrm{SP}} = 82.4722$ μm. The thicknesses of the optimized spacer layers are listed in Table S2. The transmission calculations are performed by the transfer matrix method (TMM), where at the wavelength of 4.05 μm, the refractive indices of 3.5941 and 1.3871 for poly-Si and $SiO_2$, respectively, are considered. The optimization algorithm is based on the gradient descent method that minimizes $\sigma^2$, i.e., the average over the angle range of the squared difference between the spaceplate and free-space (FS) phases, by varying the layer thicknesses.

**Table S1. Layer thicknesses of the optimized multilayered spaceplate unit structure**

| Layer number | Material | Layer thickness (μm) |
|---|---|---|
| 1 | poly-Si | 0.9828 |
| 2 | $SiO_2$ | 0.2000 |
| 3 | poly-Si | 0.9347 |
| 4 | $SiO_2$ | 0.4275 |
| 5 | poly-Si | 0.3708 |
| 6 | $SiO_2$ | 0.7215 |
| 7 | poly-Si | 1.1795 |
| 8 | $SiO_2$ | 0.5801 |
| 9 | poly-Si | 1.2922 |
| 10 | $SiO_2$ | 0.5167 |
| 11 | poly-Si | 0.8714 |
| 12 | $SiO_2$ | 0.8420 |
| 13 | poly-Si | 0.8214 |

**Table S2. Thicknesses of the optimized spacer layers associated with the coupled cascaded spaceplate structure**

| Spacer layer number | Material | Layer thickness (μm) |
|---|---|---|
| 1 | $SiO_2$ | 0.6896 |
| 2 | $SiO_2$ | 0.6446 |
| 3 | $SiO_2$ | 0.6511 |
| 4 | $SiO_2$ | 0.6425 |
| 5 | $SiO_2$ | 0.6344 |
| 6 | $SiO_2$ | 0.6426 |
| 7 | $SiO_2$ | 0.6426 |

## 2. Comparison between the transfer matrix method and commercial electromagnetic solvers

We perform transmission calculations—transmittance and the transmitted phase as a function of the incident angle—of the optimized multilayered SUS and the CCSS, with the layer thicknesses given in Tables S1 and S2, using the quasi-analytical TMM and the commercial solvers of COMSOL Multiphysics and PlanOpSim. Figure S1 shows the comparison between the TMM and the commercial solvers. Particularly, Figs. S1(a) and S1(b) show the dependence of the transmittance and the transmitted phase on the incident angle, respectively, for the optimized SUS; Figs. S1(c) and S1(d) show this dependence for the optimized CCSS. All calculations were performed for a transverse magnetic polarization. Figure S1 reveals that calculations based on the TMM and the two commercial solvers—COMSOL Multiphysics, which is based on a finite element method, and PlanOpSim, which is based on a rigorous coupled-wave analysis—exhibit excellent agreement. Quantitatively, for example, the squeezed lengths extracted from the phase calculations of the CCSS are 1.0912 mm, 1.0884 mm, and 1.0912 mm for the TMM, COMSOL Multiphysics, and PlanOpSim, respectively; these values correspond to a maximum relative error of 0.257%, which is exceedingly small. The presented comparison reflects the high accuracy of the TMM in the calculation of multilayered structures. While all methods show a similar high accuracy, the TMM is significantly faster. Therefore, we optimize the multilayered SUS and CCSS by employing the TMM.

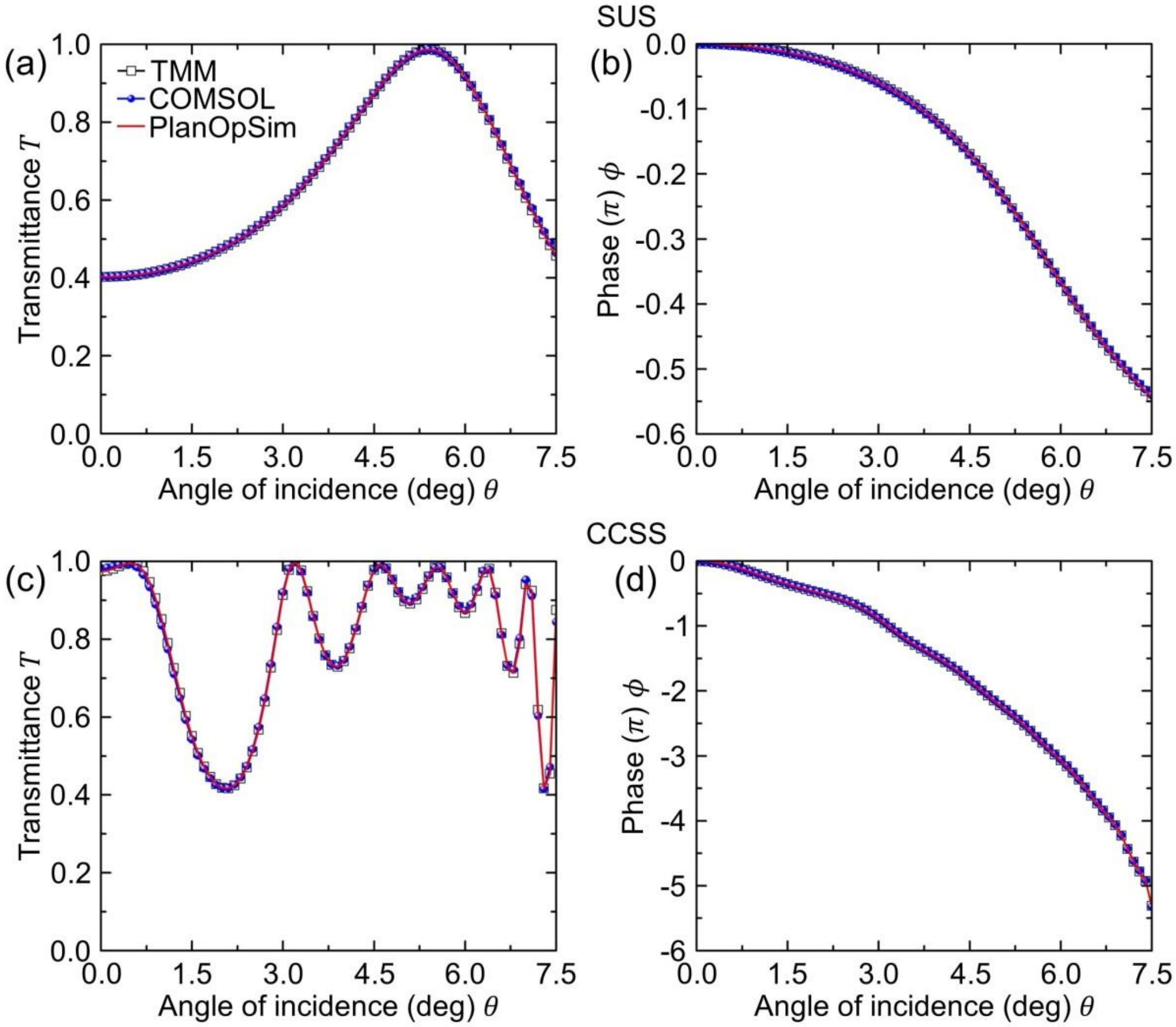

**Fig. S1.** Comparison between the transfer matrix method and commercial solvers. (a), (b) Dependence of the transmittance and phase on the angle of incidence for the optimized SUS, respectively. (c), (d) Dependence of the transmittance and phase on the angle of incidence for the optimized CCSS, respectively. The results show independent calculations using the TMM, COMSOL Multiphysics, and PlanOpSim.

## 3. Optimization method of the coupled cascaded spaceplate structure

The CCSS consists of multiple repetitions of the achieved optimized SUS, which are separated by a $SiO_2$ spacer layer. Here, unlike Fabry–Pérot resonators, the thicknesses of the spacer layers $d_{\mathrm{SL}}$ are not fixed and require further optimization. This stage is crucial, as without proper optimization of $d_{\mathrm{SL}}$, the phase of the CCSS can deviate significantly from that of the equivalent FS, resulting in a high $\sigma^2$, i.e., an undesirable low Strehl ratio $S = \exp(-\sigma^2)$. Moreover, the transmission characteristics of the CCSS are highly sensitive to the thickness of each spacer layer, so mismatched thicknesses can result in low transmittance within the target angular range. Figure S2(a) shows the Strehl ratio $S$ and the minimum transmittance $T_{\min}$ as a function of the iteration number throughout the optimization procedure of the CCSS [similar to Fig. 2(f)]. In the first stage, the uniform thicknesses initially assigned to all spacer layers are varied collectively to maximize $S$ (i.e., to minimize $\sigma^2$). In the second stage, the thickness of each spacer layer is independently fine-tuned to increase $T_{\min}$ further, while maintaining $S$ within 3% of its maximum value. This two-stage optimization procedure for the CCSS is demonstrated in Fig. S2(b), where the thickness variations of the first and second spacer layers are shown throughout the optimization—exhibiting the same thicknesses in the first stage and different thicknesses in the second stage. By optimizing the thickness of each spacer layer via this protocol, we obtain a CCSS that closely matches the equivalent FS response [see Fig. 2(d)].

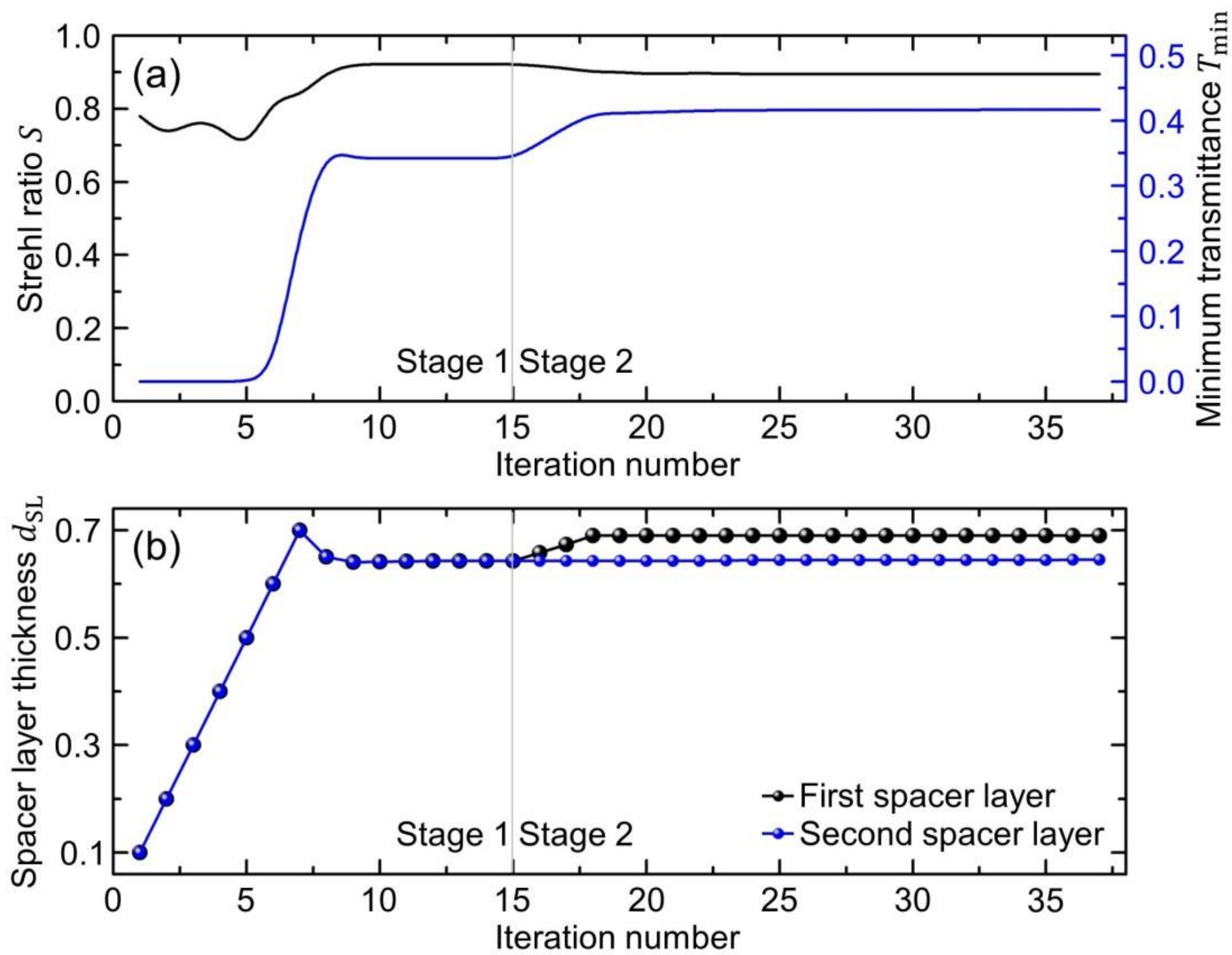

**Fig. S2.** Optimization method of the coupled cascaded spaceplate structure. (a) Strehl ratio $S$ and the minimum transmittance $T_{\min}$, in the angular range from 0° to 7.5°, as a function of the iteration number. The simultaneous convergence of $S$ and $T_{\min}$ indicates the evolution towards achieving the CCSS via the two-stage optimization. (b) Corresponding thicknesses of the first and second spacer layers throughout the optimization. The vertical line separates the iterations with uniform $d_{SL}$ (stage 1; maximizing $S$) and nonuniform $d_{SL}$ (stage 2; maximizing $T_{\min}$).